%% file: proceeding_dmueller_top2016.tex
\documentclass[12pt]{article}
\usepackage{graphicx}

\newcommand\pubnumber{ }
\newcommand\pubdate{\today}

\def\institute{Institut f\"ur Experimentelle Kernphysik\\
Karlsruhe Institute of Technology, 76131 Karlsruhe, GERMANY}

\def\Title#1{\begin{center} {\Large #1 } \end{center}}
\def\Author#1{\begin{center}{ \sc #1} \end{center}}
\def\Address#1{\begin{center}{ \it #1} \end{center}}

\newcommand\pubblock{\rightline{\begin{tabular}{l} \pubnumber\\
         \pubdate  \end{tabular}}}
\newenvironment{Abstract}{\begin{quotation}  }{\end{quotation}}
\newenvironment{Presented}{\begin{quotation} \begin{center} 
             PRESENTED AT\end{center}\bigskip 
      \begin{center}\begin{large}}{\end{large}\end{center} \end{quotation}}

\input econfmacros.tex

\begin{document}
\begin{titlepage}
\pubblock

\vfill
\Title{Investigation of Higgs boson couplings via the production of a \\single top quark in association with a Higgs boson \\in the H $\to$ b$\mathrm{\bar{b}}$ channel}
\vfill
\Author{ Denise M\"uller \\on behalf of the CMS Collaboration}
\Address{\institute}
\vfill
\begin{Abstract}
A search for the production of a single top quark in association with a Higgs boson is performed using the decay H $\to$ b$\mathrm{\bar{b}}$. The rate of this Higgs production mode is particularly sensitive to the relative sign of the Higgs boson couplings to fermions and bosons. The 2015 pp collisions data at a center-of-mass energy of 13\,TeV are analyzed. This poster focuses on the discrimination between signal and background events. Upper limits, determined in the two-dimensional plane spanned by the coupling modifiers of the Higgs boson to top quarks and to vector bosons, are presented.
\end{Abstract}
\vfill
\begin{Presented}
$9^{\mathrm{th}}$ International Workshop on Top Quark Physics\\
Olomouc, Czech Republic,  September 19--23, 2016
\end{Presented}
\vfill
\end{titlepage}
\def\thefootnote{\fnsymbol{footnote}}
\setcounter{footnote}{0}

\section{Introduction}

\begin{figure}[htb]
\centering
\includegraphics[width=0.4\textwidth]{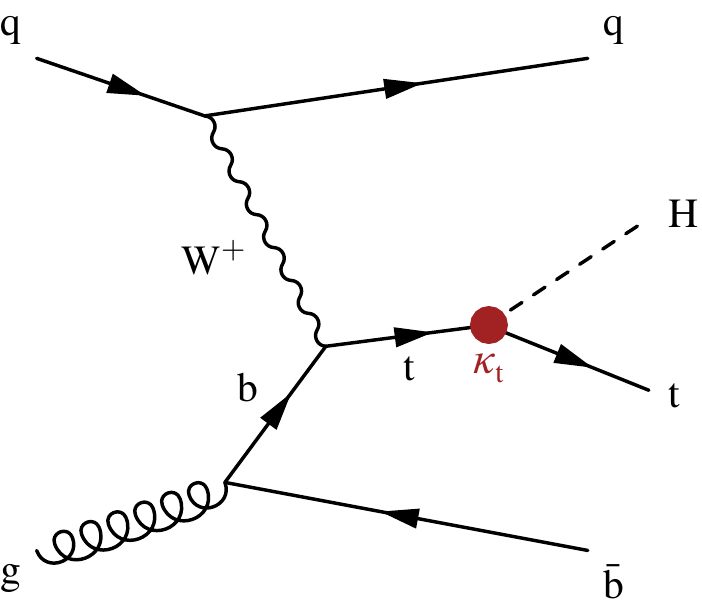}
\hspace*{1.2em}
\includegraphics[width=0.4\textwidth]{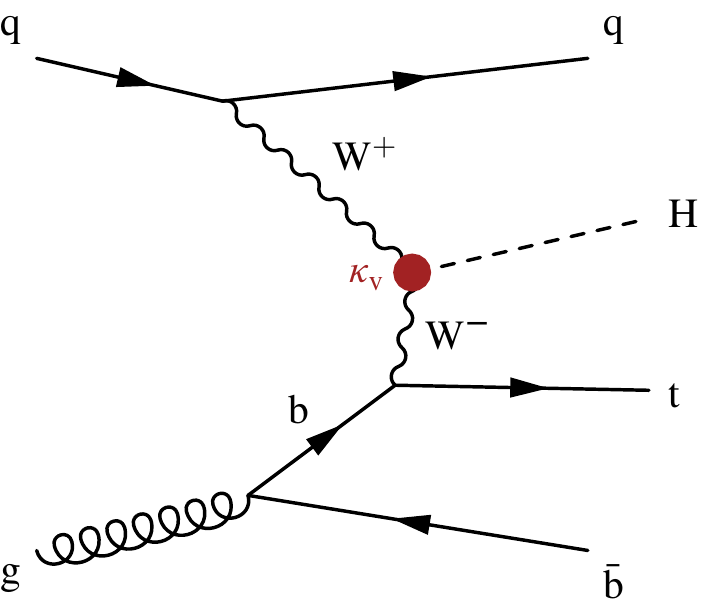}
\caption{Feynman diagrams for the associated production of a single top quark and a Higgs boson in the $t$ channel. Left figure: Higgs boson emitted from the top quark. Right figure: Higgs boson emitted from the intermediate W boson.}
\label{fig:feynman}
\end{figure}

In contrast to most other measurements, the rate of the production of a single top quark in association with a Higgs boson (tH production) is sensitive not only to the magnitude but also to the sign of the Yukawa coupling of the top quark. In this production mode, the single top quark is produced via the $t$ channel or via the associated production with a W boson. Due to its small cross section, the $s$-channel production is negligible. The Higgs boson can be emitted either from the top quark or the intermediate W boson (see Fig.~\ref{fig:feynman}) and the amplitudes of these two possibilities interfere. The resulting amplitude depends on the ratios of actual coupling strengths to the standard model (SM) predictions for the Higgs-top coupling ($\kappa_\mathrm{t}$) and for the coupling of the Higgs boson to vector bosons ($\kappa_\mathrm{V}$), given by $\mathcal{A} \propto (\kappa_\mathrm{t}-\kappa_\mathrm{V})$. A scan over different values of $\kappa_\mathrm{t}$ and $\kappa_\mathrm{V}$, suitable for a test of Higgs boson couplings, is provided. Furthermore, physics beyond the SM can be potentially discovered by the evidence of anomalous couplings. The scenario with a negative Yukawa coupling of the top quark ($\kappa_\mathrm{t} = -1.0$, $\kappa_\mathrm{V} = +1.0$) -- called inverted top coupling (ITC) scenario -- is particularly relevant for the analysis, which differs only in the sign of $\kappa_\mathrm{t}$ from the SM case ($\kappa_\mathrm{t} = +1.0$, $\kappa_\mathrm{V} = +1.0$), as it provides a significantly higher cross section. The increase of the cross section is caused by the constructive interference of the amplitudes of the two emission possibilities of the Higgs boson. In the SM case, a destructive interference of the amplitudes occurs.

\section{Classification and Limit Calculation}

The analyzed dataset consists of events recorded with the Compact Muon Solenoid (CMS) experiment~\cite{det} during Run II of the Large Hadron Collider (LHC) in 2015, corresponding to an integrated luminosity of $2.3\,\mathrm{fb}^{-1}$. Events are selected which contain exactly one isolated lepton (muon or electron), three or four b-tagged jets and at least one untagged jet. All jets are required to have $p_\mathrm{T}>30\,\mathrm{GeV}$ ($|\eta|<2.4$) and $p_\mathrm{T}>40\,\mathrm{GeV}$ ($|\eta|\geq 2.4$) respectively. Additionally, a cut for the missing transverse energy is applied: ${\not\mathrel{E}}_\mathrm{T}>45\,\mathrm{GeV}$ (electron channel) and ${\not\mathrel{E}}_\mathrm{T}>35\,\mathrm{GeV}$ (muon channel). According to the number of b-tagged jets, two independent signal regions are defined, namely the 3 tag and 4 tag region.

For the event classification, 51 boosted decision trees (BDTs) are used. Each of these BDTs corresponds to one point in the two-dimensional $\kappa_\mathrm{t}-\kappa_\mathrm{V}$ plane which consists of 51 different values ranging from $-3.0 \leq \kappa_\mathrm{t} \leq +3.0$ and $\kappa_\mathrm{V} = +0.5, +1.0, +1.5$. For all BDTs, the signal events (tH production) are trained against the dominating t$\mathrm{\bar{t}}$ and t$\mathrm{\bar{t}}$H background in the 3 tag region. Three types of input variables are used for the classification BDTs: Variables from one of the 51 tH reconstructions, variables from the t$\mathrm{\bar{t}}$ reconstruction and global variables, which are independent of any reconstruction. The most discriminating variables are $\log m(\mathrm{t}_\mathrm{had})$ (from t$\mathrm{\bar{t}}$ reconstruction), $|\eta(\mathrm{recoil\ jet})|$ (from tH reconstruction) and aplanarity (global variable). The outputs of the classification BDTs are applied to the 3 tag and to the 4 tag region.

After the event classification, limits for 51 different points in the $\kappa_\mathrm{t}-\kappa_\mathrm{V}$ plane are determined from a simultaneous fit of the corresponding BDT output in the 3 tag and 4 tag region. The largest systematic uncertainty of the limit calculation arises from variations in the jet energy scale, from variations in the $Q^2$ scale used in the generation of the t$\mathrm{\bar{t}}$ and tH samples, and from b-tagging reweighting. 

\section{Results}

\begin{figure}[h]
\centering
\includegraphics[width=0.45\textwidth]{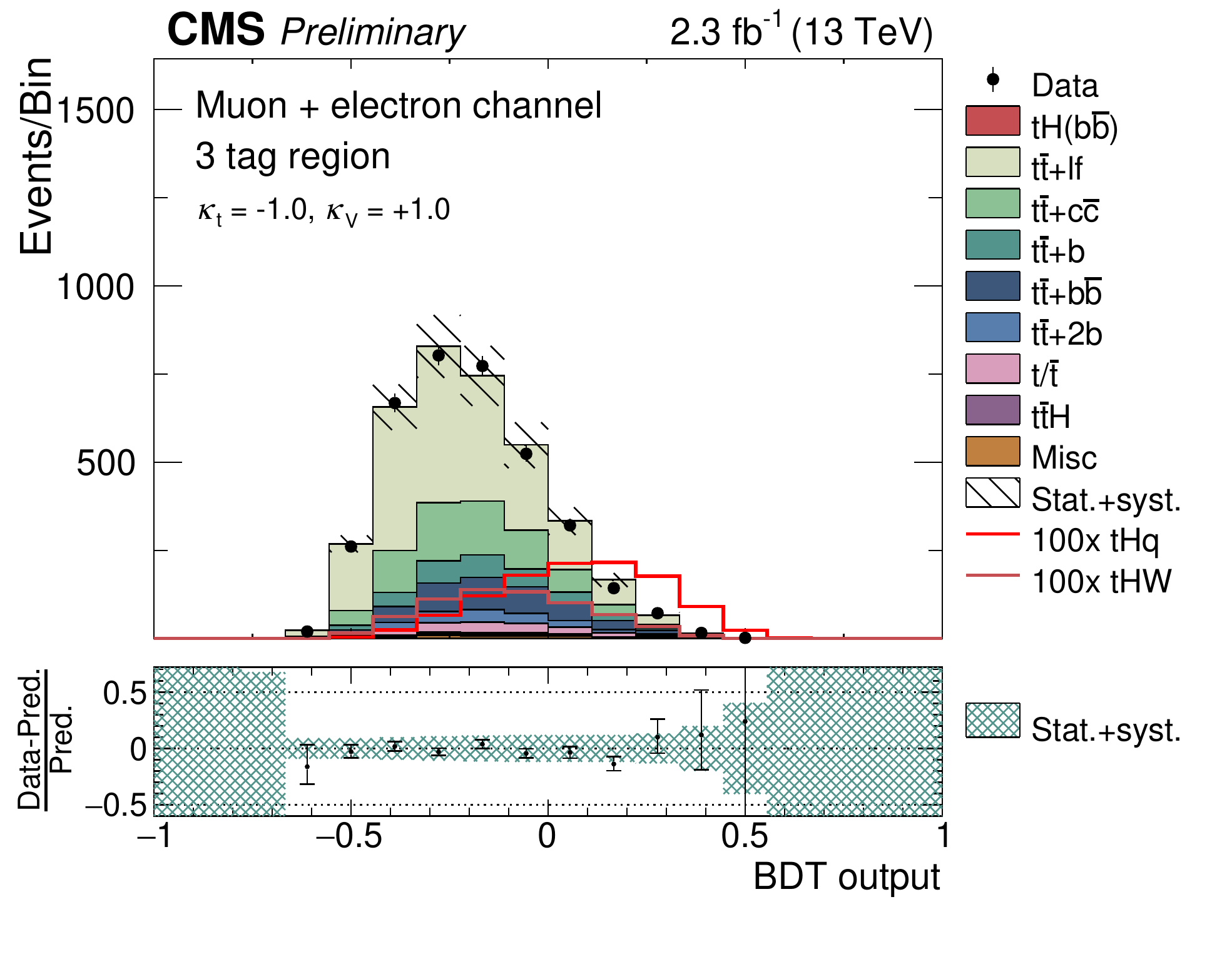}
\hspace*{1.2em}
\includegraphics[width=0.45\textwidth]{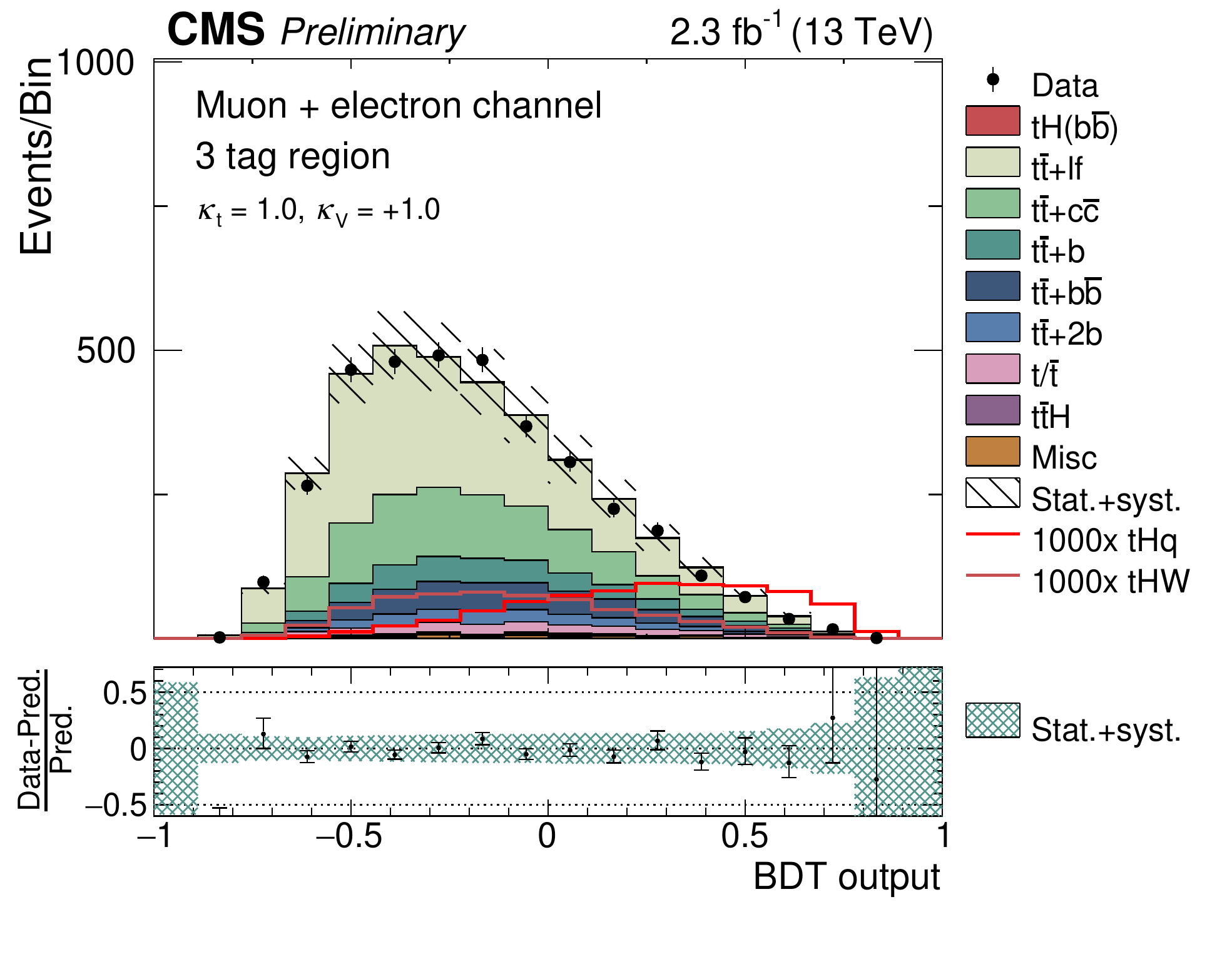}
\caption{Postfit distributions of the classification BDT output in the 3 tag region for the ITC (left) and SM (right) scenario. The signal distributions correspond to the expected contributions scaled by the factors given in the legends. Taken from~\cite{pas}.}
\label{fig:postfit-BDT1}
\end{figure}

\begin{figure}[h]
\centering
\includegraphics[width=0.45\textwidth]{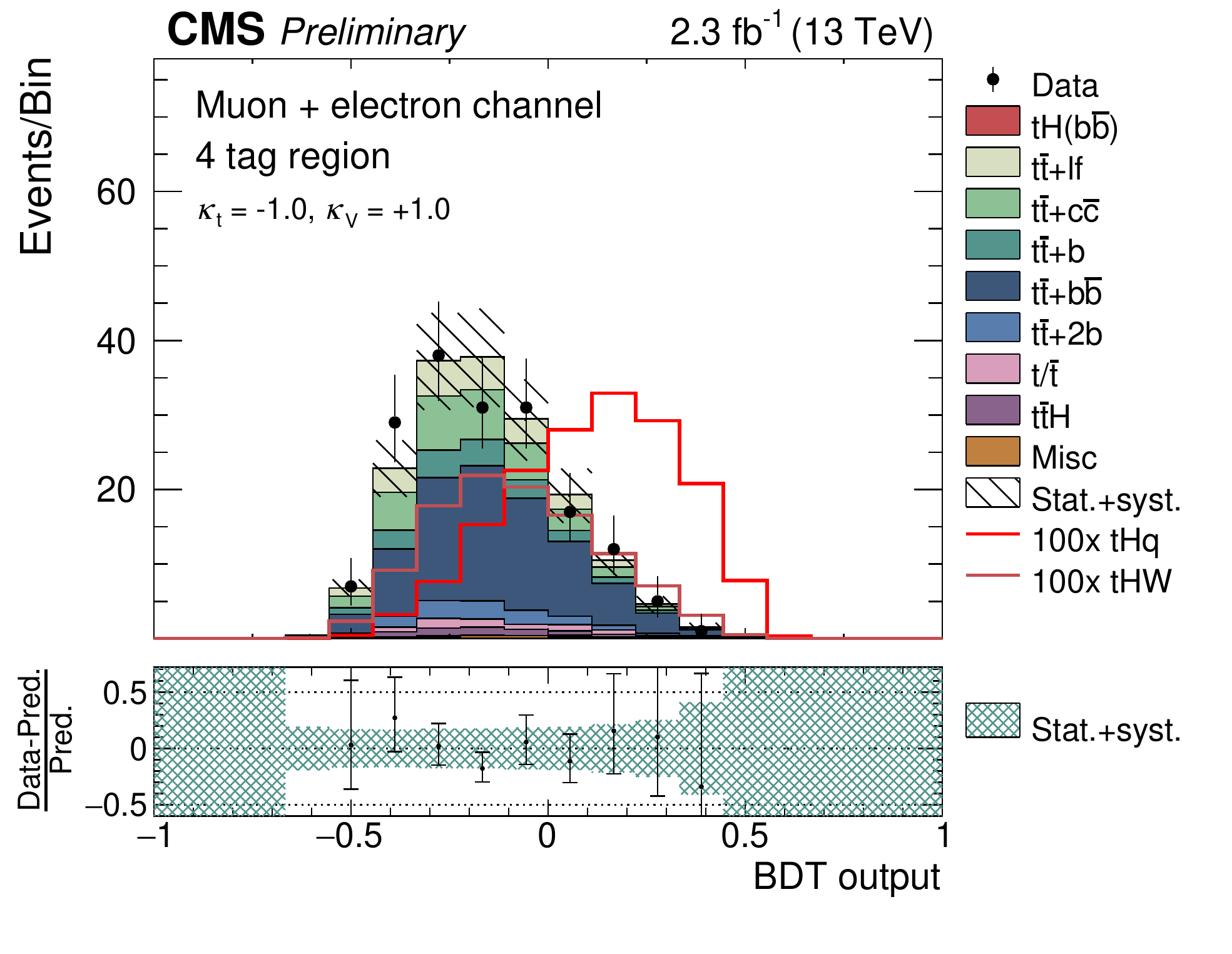}
\hspace*{1.2em}
\includegraphics[width=0.45\textwidth]{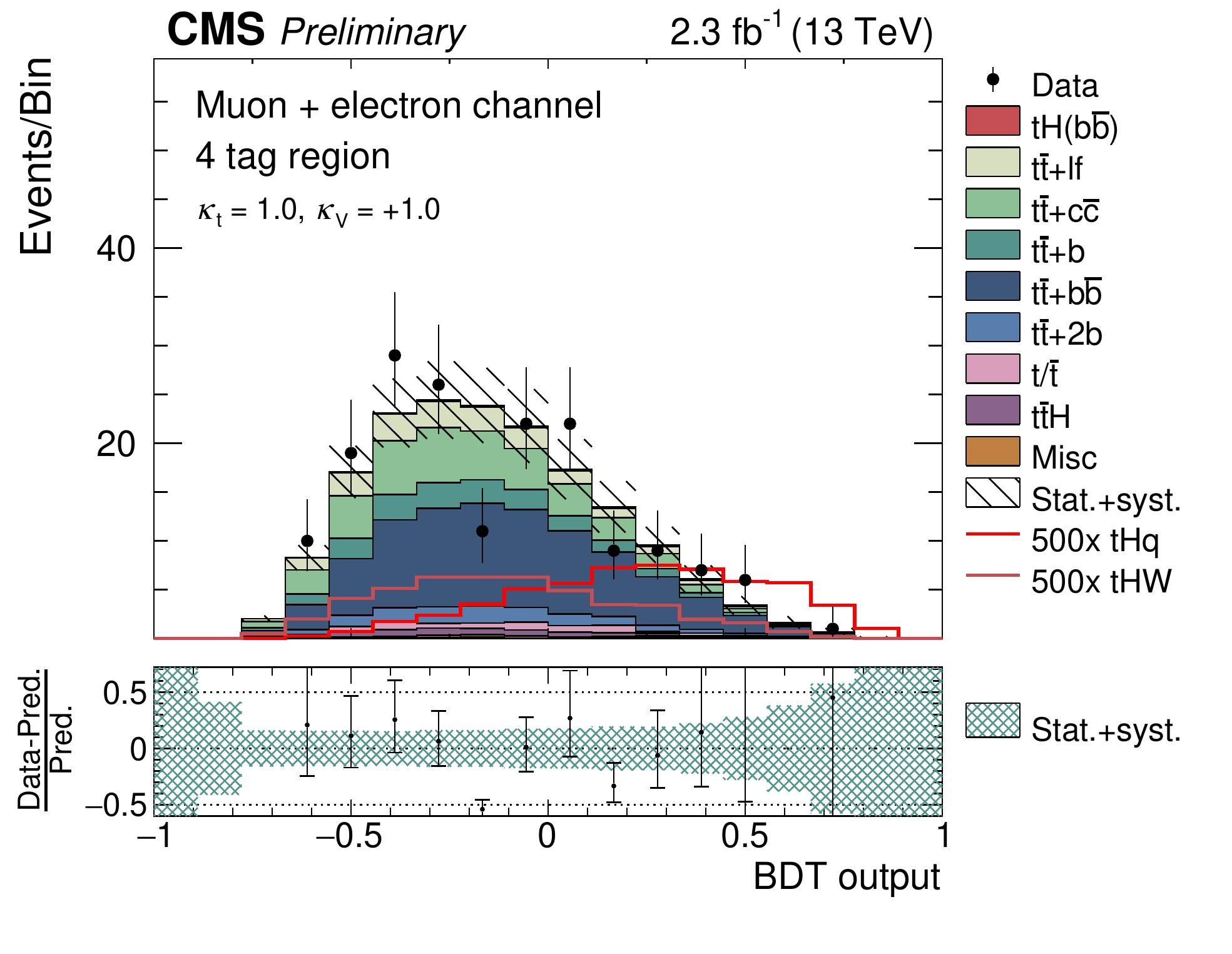}
\caption{Postfit distributions of the classification BDT output in the 4 tag region for the ITC (left) and SM (right) scenario. The signal distributions correspond to the expected contributions scaled by the factors given in the legends. Taken from~\cite{pas}.}
\label{fig:postfit-BDT2}
\end{figure}

\begin{figure}[h!]
\centering
\includegraphics[width=0.45\textwidth]{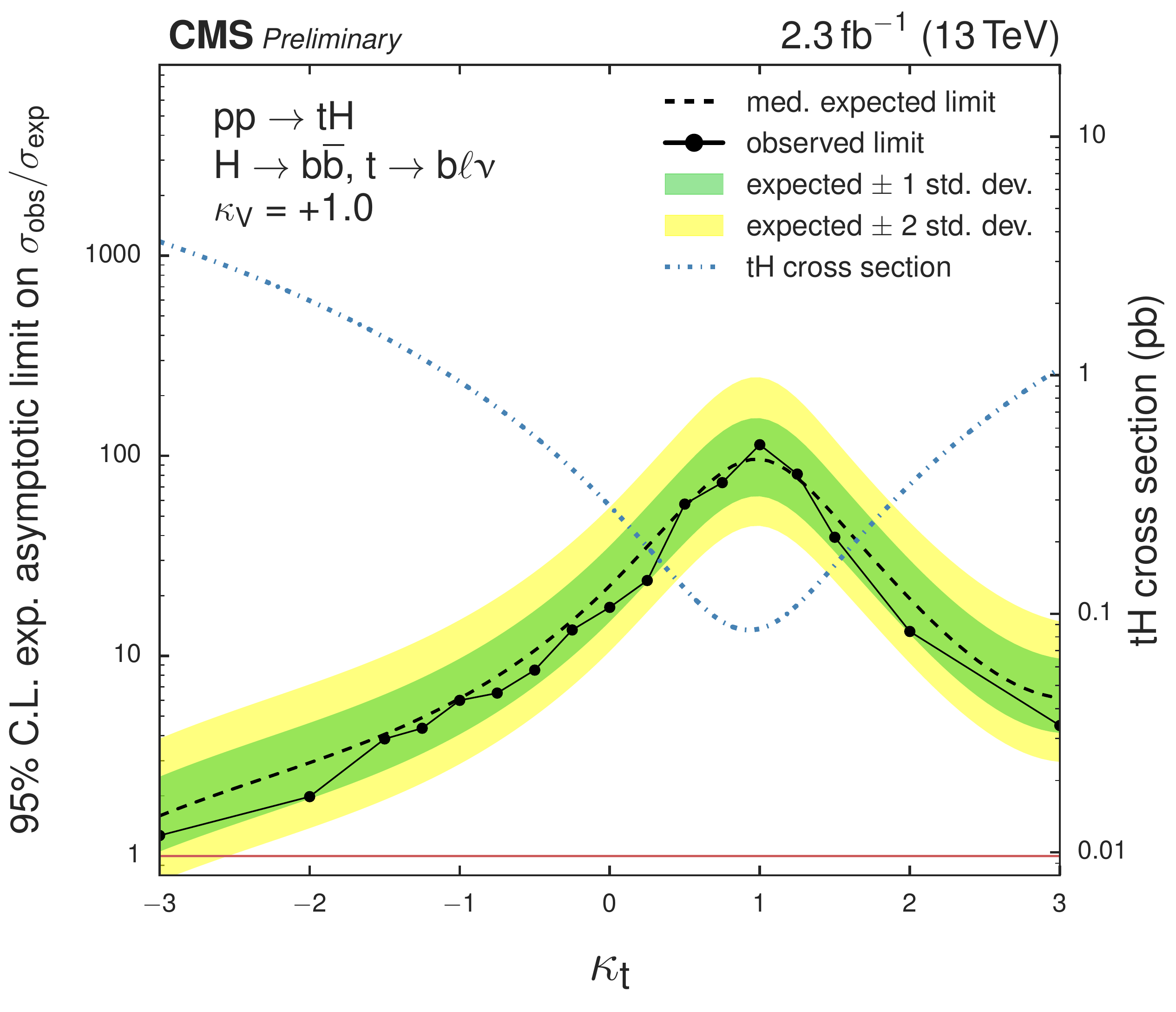}\\
\includegraphics[width=0.4\textwidth]{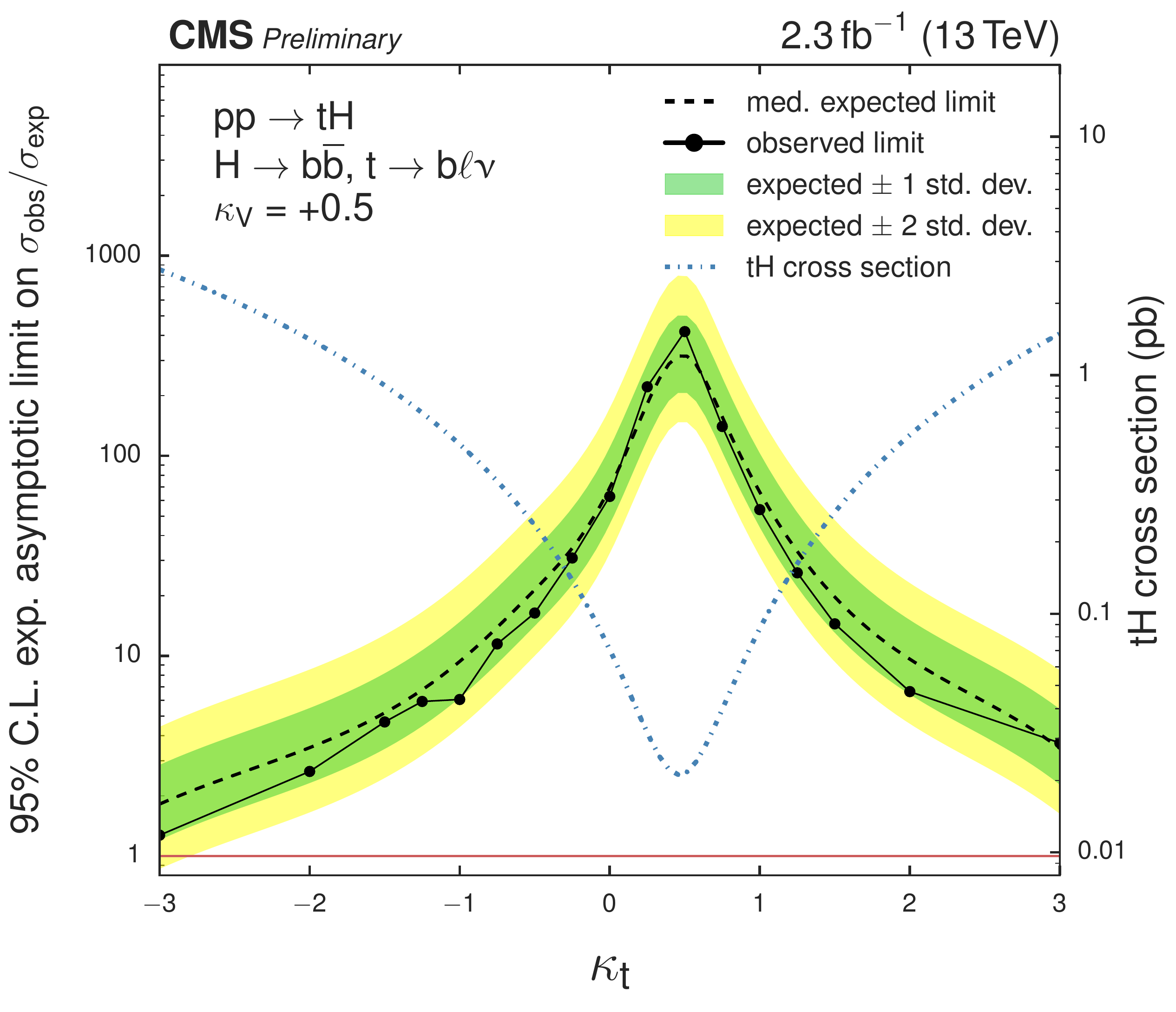}
\includegraphics[width=0.4\textwidth]{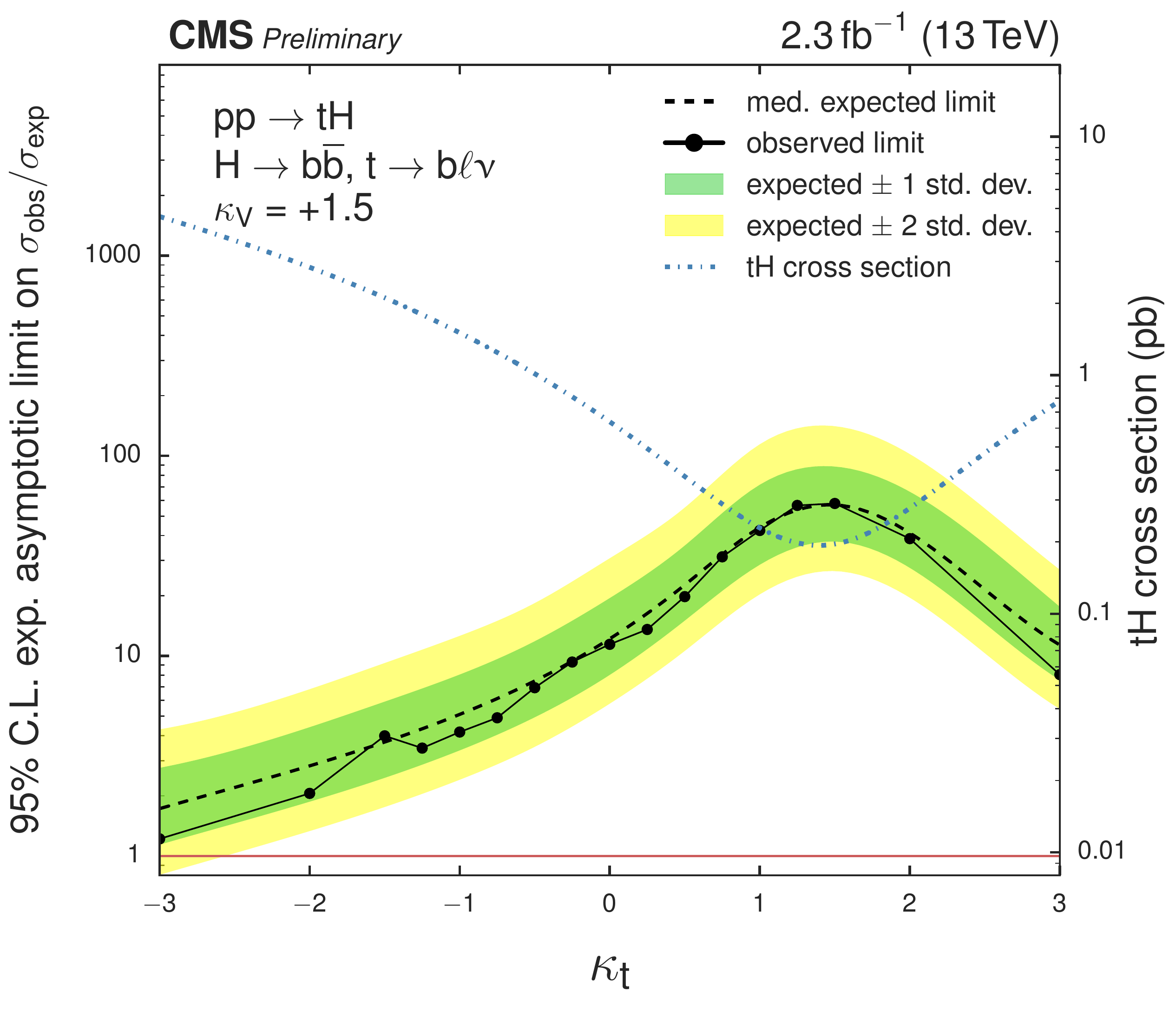}
\caption{Upper limits on tH scenarios with different $\kappa_\mathrm{t}-\kappa_\mathrm{V}$ configurations. Top figure: $\kappa_\mathrm{V}=+1.0$, bottom left figure: $\kappa_\mathrm{V}=+0.5$, bottom right figure: $\kappa_\mathrm{V}=+1.5$. The tH cross sections are given on the right $y$ axis. Taken from~\cite{pas}.}
\label{fig:limits}
\end{figure}

The resulting postfit distributions of the classification BDT output for the ITC and for the SM coupling scenario in the two signal regions are shown in Fig.~\ref{fig:postfit-BDT1} and \ref{fig:postfit-BDT2}. The expected and observed upper limits on the tH production rate for all 51 studied couplings can be found in Fig.~\ref{fig:limits}. For the SM case, the observed limit is $113.7\times\sigma_\mathrm{SM}$ (expected: 98.6), and for the ITC case, an upper limit of $6.0\times\sigma_\mathrm{ITC}$ is observed (expected: 6.4). For all 51 points in the $\kappa_\mathrm{t}-\kappa_\mathrm{V}$ plane, the observed limit is well within one standard deviation of the expected limit. The sensitivity is already comparable to the Run I analysis~\cite{pas_old} which yielded an expected limit of $5.4\times\sigma_\mathrm{ITC}$ for the ITC scenario.

A more detailed description of the analysis is available in Ref.~\cite{pas}.

\clearpage

\end{document}

%% file: econfmacros.tex
\def\beq{\begin{equation}}
\def\eeq#1{\label{#1}\end{equation}}
\def\eeqn{\end{equation}}

\def\beqa{\begin{eqnarray}}
\def\eeqa#1{\label{#1}\end{eqnarray}}
\def\eeqan{\end{eqnarray}}

\let\bar=\overbar

\def\Dslash{\not{\hbox{\kern-4pt $D$}}}
\def\dslash{\not{\hbox{\kern-2pt $\del$}}}

\def\msb{{\bar{\ssstyle M \kern -1pt S}}}

